\documentclass[iop]{emulateapj}

\usepackage{natbib,graphicx,amsmath,xspace}

\newcommand{\fluxu}{\mbox{ergs cm$^{-2}$ s$^{-1}$}\ }
\newcommand{\mcg}{\mbox{MCG--5-23-16}\xspace}
\newcommand{\ngc}{\mbox{NGC 7314}\xspace}
\newcommand{\rg}{\mbox{$r_g$}\xspace}

\begin{document}

\title{Discovery of Fe K$\alpha$ X-ray reverberation around the black holes in \mcg and \ngc}

\author{A. Zoghbi$^{1,2}$, C. Reynolds$^{1,2}$, E. M. Cackett$^3$, G. Miniutti$^4$, E. Kara$^5$, A. C. Fabian$^5$}
\affil{$^1$Department of Astronomy, University of Maryland, College Park, MD 20742-2421, USA}
\affil{$^2$Joint Space-Science Institute (JSI), College Park, MD 20742-2421, USA}
\affil{$^3$Department of Physics \& Astronomy, Wayne State University, 666 W. Hancock St, Detroit, MI 48201}
\affil{$^4$Centro de Astrobiologia (CSIC-INTA), Dep. de Astroﬁsica; PO Box 78, E-28691, Villanueva de la C\~{a}nada, Madrid, Spain}
\affil{$^5$Institute of Astronomy, Madingley Road, Cambridge CB3 0HA. UK}
\email{azoghbi@astro.umd.edu}

\begin{abstract}
Several X-ray observations have recently revealed the presence of reverberation time delays between spectral components in AGN. Most of the observed lags are between the power-law Comptonization component, seen directly, and the soft excess produced by reflection in the vicinity of the black hole. NGC 4151 was the first object to show these lags in the iron K band. Here, we report the discovery of reverberation lags in the Fe K band in two other sources: \mcg and \ngc. In both objects, the $6-7$ keV band, where the Fe K$\alpha$ line peaks, lags the bands at lower and higher energies with a time delay of $\sim1$ kilo-seconds. These lags are unlikely to be due to the narrow Fe K$\alpha$ line. They are fully consistent with reverberation of the relativistically-broadened iron K$\alpha$ line. The measured lags, their time-scale and spectral modeling, indicate that most of the radiation is emitted at $\sim$5 and 24 gravitational radii for \mcg and \ngc respectively.
\end{abstract}

\section{Introduction}
Observations of active galactic nuclei (AGN) have shown that most of the X-ray radiation is emitted very close to the central super-massive black hole.
Short time scale variability (e.g. \citealt{2011MNRAS.413.2489V}), relativistic spectral distortions (\citealt{1995Natur.375..659T, 2003PhR...377..389R,2007ARA&A..45..441M, 2012MNRAS.419..116F}), X-ray eclipses (\citealt{2007ApJ...659L.111R}), as well as gravitational micro-lensing measurements (\citealt{2012ApJ...757..137C}), all indicate that emission originates in a compact region a few gravitational radii from the event horizon.

The primary X-ray radiation is thought to be produced in a hot electron corona by Compton up-scattering lower energy disk photons (\citealt{1991ApJ...380L..51H}). This radiation illuminates the surrounding, relatively cold, matter (\citealt{1988MNRAS.233..475G, 1989MNRAS.238..729F}) giving rise to characteristic reflection spectra (\citealt{1993MNRAS.261...74R}). The variability of the primary X-ray source in this environment naturally leads to time delays comparable to the light-crossing time to the reflector (\citealt{1999ApJ...514..164R}). The first direct evidence for such delays was measured in the Narrow Line Seyfert 1 (NLS1) galaxy 1H0707-495 (\citealt{2009Natur.459..540F, 2010MNRAS.401.2419Z}). It was observed as a $\sim30$ seconds delay between the primary X-ray source and the strong iron L line produced by reflection. Several subsequent studies have shown that such lags are present in many other Seyfert galaxies possessing a soft excess (\citealt{2011MNRAS.416L..94E, 2011MNRAS.417L..98D, 2011MNRAS.418.2642Z, 2013MNRAS.428.2795K, 2012arXiv1208.5898F, 2012arXiv1210.7874C}). All these delays are comparable to the light-crossing time at a few gravitational radii (\rg) from the black hole, and there is even evidence that the amplitude and frequency of the lag correlates with the black hole mass (\citealt{2012arXiv1201.0196D}).

A key property in all those objects is their strong soft excess, a characteristic feature in the NLS1 class. The measured time delays between the direct power-law component, that dominates the $2-4$ keV band, and the soft excess ($<1$ keV) are naturally explained if the latter is due to a reflection process. In this case, a combination of emission lines and bremsstrahlung from the surface layers of an ionized disk are smoothed and distorted by relativistic effects. Interpreting the measured lags as light-crossing effects in the immediate vicinity of the black hole was questioned by \cite{2010MNRAS.408.1928M} who argued that these delays could be an artifact of a much larger system (hundreds of \rg). Although such an interpretation fails when a full set of observed timing and spectroscopic properties are considered (\citealt{2011MNRAS.412...59Z}), it partly stands on the ambiguity of interpreting the soft excess itself. The spectrum of the observed soft excess tend to be very smooth masking out any direct detection of spectral emission features.  Detecting reverberation of the broad iron K-alpha line at $\sim 6$\,keV would remove such ambiguity and clearly demonstrate an accretion disk origin for the time lags.

Lags between energy bands containing the iron-K line and softer energy bands have been known about for some time, both in Galactic black holes (\citealt{1988Natur.336..450M, 2001MNRAS.327..799K}) and AGN (\citealt{2007MNRAS.382..985M,2008MNRAS.388..211A}).
They are seen at low frequencies (typically around $10^{-4}$ Hz for a $10^6$ M$_{\odot}$ black hole, and lower frequencies for higher masses or larger emission radius, e.g. \citealt{2011MNRAS.412...59Z,2013MNRAS.tmp..654K,2012arXiv1201.0196D}) and show an almost featureless energy spectrum (with $\Delta t\sim \log E$); which, in analogy with Galactic black holes (where the lags extend well above 10 keV), leads most authors to associate them with continuum process itself (e.g. Comptonization or propagating fluctuations. \citealt{2001MNRAS.327..799K, 2001AdSpR..28..267P,2007MNRAS.382..985M,2011MNRAS.412...59Z}). On the other hand, \cite{2010MNRAS.408.1928M, 2010MNRAS.403..196M} attribute such low-frequency lags to reverberation associated with the continuum scattering in a hypothesized extended structure.  A major recent development was the discovery that the lags in NGC~4151 which has an energy dependence that traces the shape of a relativistically-broad iron line (\citealt{2012MNRAS.422..129Z}). This is a clear and dramatic confirmation of a central prediction of the relativistic disk-reflection paradigm. The observed lag in this case is an extension of the {\it soft} reverberation lags seen in many objects (\citealt{2009Natur.459..540F, 2010MNRAS.401.2419Z} and later work), and it is separate from the low-frequency featureless continuum lag.

In addition, the `line' seen in the lag-energy spectrum of NGC~4151 has a stronger blue horn at long time-scales and a prominent red wing at short time-scales, in a striking match to the expectation of a line emitted from a disk and distorted by relativistic effects (\citealt{1989MNRAS.238..729F}). Detection of Fe K lags has also been seen in 1H0707-495 (\citealt{2013MNRAS.428.2795K}) and IRAS 13224-3809 (\citealt{2013MNRAS.428.2795K}). Here, we report on the detection of similar lags in two Seyfert galaxies \mcg and \ngc using XMM-Newton observations. The two objects were selected because of their brightness and high variability in the $2-10$ keV band. Their selection came during the search for the best reverberation targets for the proposed X-ray mission GRAVITAS (\citealt{2012ExA....34..445N}).

\mcg is a Seyfert 1.9 Galaxy ($z=0.0085$) with a typical 2--10 keV flux of $\sim8\times10^{-11}$ \fluxu and a mass of $5\times10^7$ M$_{\odot}$ (\citealt{1986ApJ...306L..61W}). Its spectrum resembles a classical Compton-thin Seyfert 2 galaxy with a column that does not affect the spectrum above 3 keV significantly. The spectrum below 1 keV contains a combination of emission from scattered continuum photons and distant photoionized gas. Above 2 keV, the spectrum shows both a narrow (EW$\sim$60 eV) and broad (EW$\sim50-200$ eV) iron K$\alpha$ lines along with a strong Compton hump above 10 keV (\citealt{1997ApJ...474..675W, 2004ApJ...601..771M, 2006AN....327.1067B, 2007PASJ...59S.301R}).

\ngc is also classified as a Seyfert 1.9 galaxy ($z=0.0048$) with an estimated mass of $5\times10^6$ M$_{\odot}$ (\citealt{1994A&A...288..425S}). It shows strong variability on all observed time scales. It is thought to be a type 2 counterpart to the NLS1 class (\citealt{2005ApJ...625L..31D}). The source is possibly seen through a weak warm absorber. The iron K band also shows both a narrow and broad components in observations with ASCA, Chandra and XMM (\citealt{1996ApJ...470L..27Y, 2003ApJ...596...85Y, 2011A&A...535A..62E}). It has also been shown that the narrow and broad components of the line respond differently to variations in the continuum (\citealt{ 2003ApJ...596...85Y}). Both objects show hard and absorbed spectra. We concentrate in our analysis on energies $>2$ keV to avoid any spectral complexities due to either emission from scattered and distant continuum photons and photoionized gas, or from warm absorption.

The paper is organized as follows: Section \ref{sec:data} describes the data used in the analysis. The results for \mcg and \ngc are presented in Sections \ref{sec:mcg} and \ref{sec:ngc} respectively, which include both lag and spectral modeling. The interpretation and implications of the results are presented in \mbox{Section \ref{sec:disc}}.

\begin{figure}[t]
\includegraphics[width=0.4\textwidth,clip]{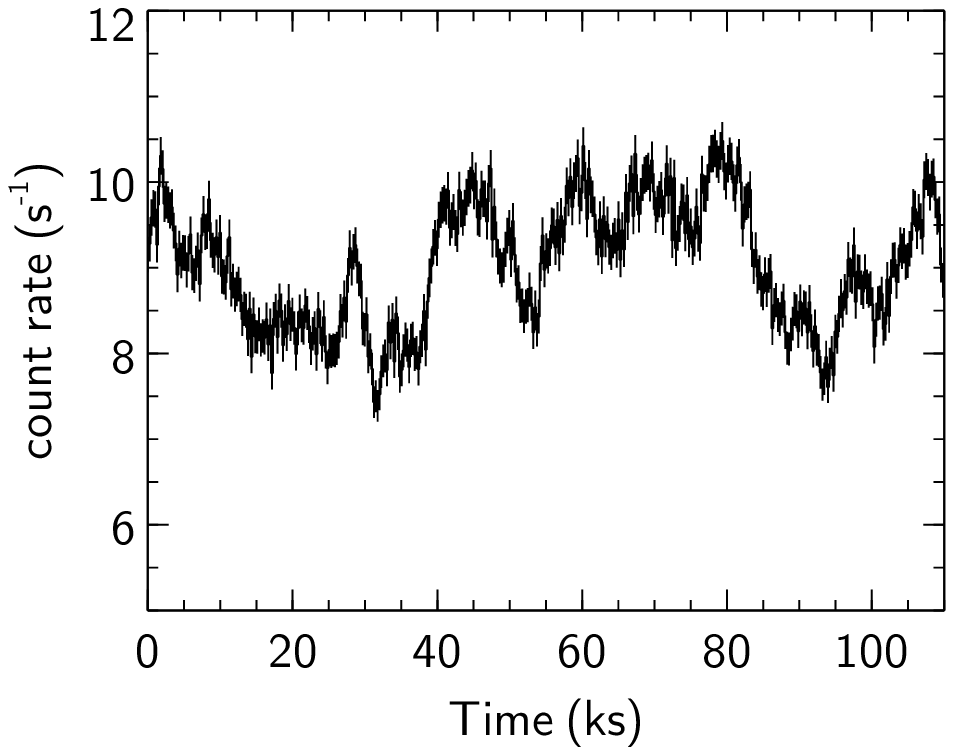}
\caption{The light curve of \mcg in the $2-10$ energy band. It shows the high count rate and strong variability.}
\label{fig:mcg_lc}
\end{figure}

\section{Data Analysis}\label{sec:data}
The results presented here are obtained using archival XMM observations of \mcg and \ngc. \mcg was observed three times (Obs ids: 0112830301, 0112830401 and 0302850201). The first had a strong particle background and is not used. The second observation is short (20 ks after high background filtering) and not very useful for timing analysis. Only the third long observation was used for timing (exposure: 130 ks) after removing the last $\sim20$ ks affected by strong particle background in the detector. This observation was taken in the large window mode and had some pileup, with a deviation from the model by 1.6 and 3.7 percent for the single and double patterns respectively (as measured by {\tt epatplot} in SAS). Pile-up is not an issue for timing analysis as show by \cite{2011MNRAS.418.2642Z}, who performed detailed simulations on  the effect of pile-up on lag measurements. The central regions were exercised in the spectral modeling to remove the most affected regions in the psf. \ngc was observed only once (obs id: 0111790101; exposure: 44 ks) in a small window mode with an additional one off-axis observation (obs id: 0311190101; exposure 83 ks) and both observations were used. The ODF files were reduced with the standard pipeline using SAS 12.0.1. Source and background light curves were extracted from circular regions of 50 arcsec radii from the pn detector. The light curves were then corrected for instrumental effects using {\tt epiclccorr}. Spectra and responses were extracted following the standard procedure in the XMM ABC guide.

\begin{figure}[t]
\includegraphics[width=0.37\textwidth,clip]{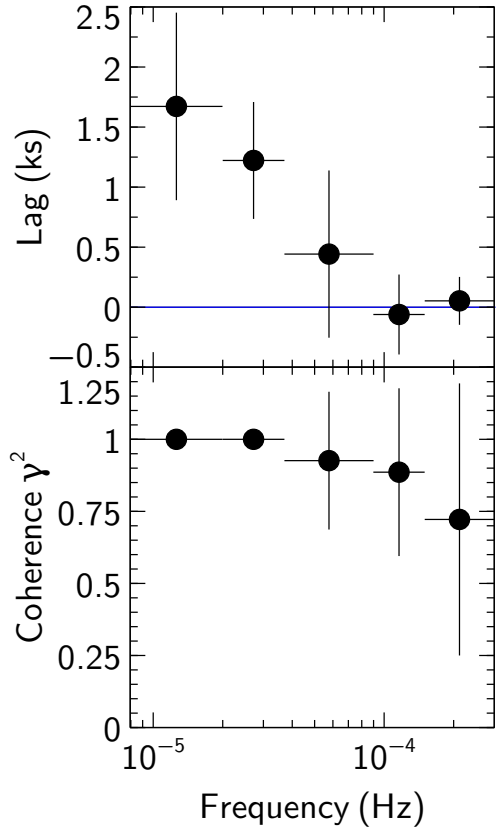}
\caption{The lag (top) and coherence (bottom) as a function of Fourier frequency between $4-5$ and $6-7$ keV energy bands for \mcg. Positive lags in the top panel indicate hard lags, where the harder band ($6-7$ keV, where the reflected iron K line peaks) lags the softer band ($4-5$ keV, dominated by the direct emission). There is a significant hard lag at frequencies $<10^{-4}$ Hz. The measured coherence remains high at those frequencies.}
\label{fig:mcg_lagcoh}
\end{figure}

\section{\mcg}\label{sec:mcg}
\subsection{Time lags}\label{sec:mcg_lag}
Fig. \ref{fig:mcg_lc} shows the $2-10$ keV light curve. The object is clearly very bright and variable. To study time lags, light curves were constructed in eight energy bins between 2 and 10 keV with 1 keV bin width. We used standard Fourier techniques to calculate the frequency-dependent time lags (\citealt{1999ApJ...510..874N}) and found that the lag depends on both frequency and on the energy bands used. As an example Fig. \ref{fig:mcg_lagcoh} (Top) shows the frequency-dependent lag between the $4-5$ and $6-7$ keV energy bands. We use the standard sign convention, where positive values indicate a hard lag (i.e. the harder band is delayed with respect to the softer band). It is clear that the $6-7$ keV band lags behind the $4-5$ keV band below $10^{-4}$ Hz. The bottom panel of Fig. \ref{fig:mcg_lagcoh} shows the measured coherence function $\gamma^2$ (\citealt{1997ApJ...474L..43V}). This is a measure of the fraction of one light curve that can be predicted from the other. It can be seen that the coherence is high and the light curves match each others variability.

\begin{figure}[t]
\includegraphics[width=0.4\textwidth,clip]{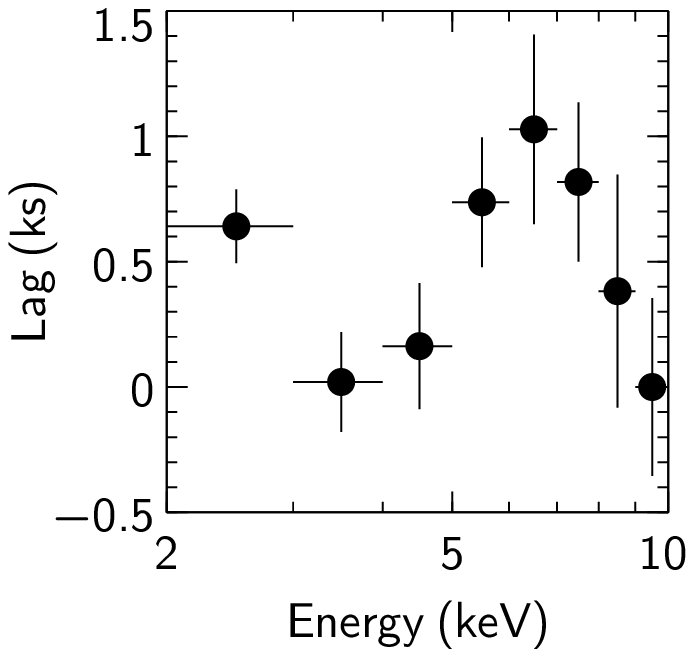}
\caption{The lag as a function of energy for frequencies $<10^{-4}$ for \mcg. Energy bins have a width of 1 keV. The lags are measured with respect to a reference band taken to be the whole $2-10$ keV band (see text for details).}
\label{fig:mcg_lagen2}
\end{figure}

In order to study the energy-dependence of the lag, we select the frequency band where there is a positive lag in Fig. \ref{fig:mcg_lagcoh}-top, i.e. $f<10^{-4}$ Hz, and calculate the lag for the different energy bands with respect to a reference band. The reference can be any band, or can be the total $2-10$ keV band itself. Here, for each energy band, the reference band is taken to be the whole $2-10$ keV band excluding the current band. Using the whole band as a reference maximizes the signal to noise, while removing the band of interest from the reference ensures that the noise remains uncorrelated (see \citealt{2011MNRAS.412...59Z}). The result is a plot of the lag at frequencies $<10^{-4}$ Hz as a function of energy, and it is shown in Fig. \ref{fig:mcg_lagen2}.

%Note that the zero point of the y-axis in Fig. \ref{fig:mcg_lagen2} is arbitrary. It depends on the reference band, and because the reference is taken to be the whole $2-10$ keV band, the zero point is the average value of the lags. If we were to choose for example the $3-4$ keV band as a reference, then the whole plot shifts towards positive lags by $\sim0.5$ ks.

\begin{figure}[ht]
\includegraphics[width=0.46\textwidth,clip]{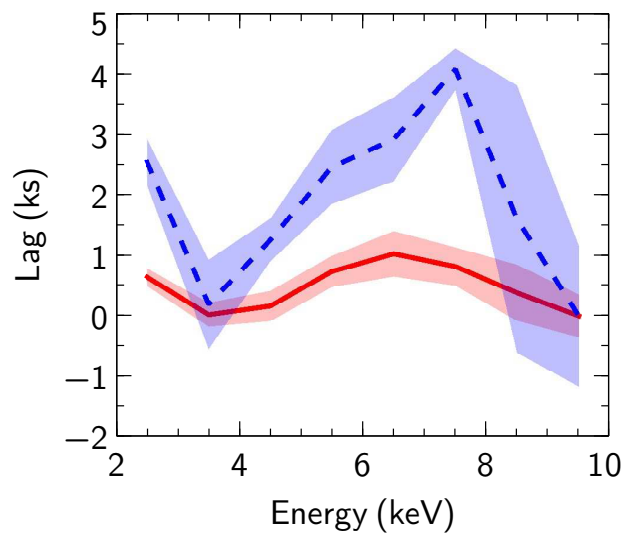}
\caption{The lag as a function of energy for frequencies $<10^{-4}$ (red; continuous line) and $<2\times 10^{-5}$ (blue; dashed line) for \mcg. The probability that the two lines have different peaks is $85\%$.}
\label{fig:mcg_lagen}
\end{figure}

Fig. \ref{fig:mcg_lagen2} clearly shows that the lag traces the shape of a line that peaks in the $6-7$ keV band, where the iron K$\alpha$ line peaks. This is similar to the shape reported for NGC 4151 (\citealt{2012MNRAS.422..129Z}), but appears to be less broad.
Note that the zero point in Fig. \ref{fig:mcg_lagen2} (and the rest of the lag-energy plots in this work) depends on the chosen reference band, but for clarity, we have shifted the plot vertically so that the zero point is the point with the most negative lag (the last point in this case).
Note also that the average value of the positive lag at frequencies $<10^{-4}$ Hz in Fig. \ref{fig:mcg_lagcoh}-top is $\sim 0.7$ ks, which is the value of the vertical difference between the $4-5$ and $6-7$ keV points in Fig. \ref{fig:mcg_lagen2}.
To look for frequency-dependent change in the feature at $\sim 6$ keV, we also produced the same plot for the lowest frequencies ($<2\times10^{-5}$). Fig. \ref{fig:mcg_lagcoh}-top indicate that the lags becomes larger as we move to lower frequencies, and this is indeed what we see in Fig. \ref{fig:mcg_lagen}.

The figure shows the lag for both frequency bands $<10^{-4}$ Hz (red continuous line) and $<2\times10^{-5}$ Hz (blue dashed line). The lag magnitude increases at lower frequencies. There are even hints of a possible change of shape of the broad line, although not with a high significance. The probability of having different peak energies is $\sim 85\%$ when fitted with gaussians with different widths.

\begin{deluxetable}{ccc}
\tablecolumns{3}
\tablewidth{0pc}
\tablecaption{Best fit parameters.\label{tab:fits}}
\tablehead{ \colhead{Parameter}    &  \colhead{\mcg} &   \colhead{\ngc} }
\startdata
Absorbing $n_h$		&	$1.59\pm0.06$			&	$0.44\pm0.13$		\\
$\Gamma$			&	$1.70^{+0.03}_{-0.13}$	&	$1.66\pm0.05$		\\
$r_{\rm in}$(\rg)	&	$5.6^{+13.1}_{-4.3}$		&	$24^{39}_{15}$		\\
Fe abundance (Fixed)&	$1$						&	$1$					\\
emm. index $q$		&	$2.2\pm0.5$				&	$2.1^{+0.5}_{-1.4}$	\\
Inclination angle($^{\circ}$)	&	$38^{+9}_{-8}$			&	$9^{+10}_{-9}$		\\
Gaussian Line		&	\nodata					&	$6.92^{+0.09}_{-0.06}$ keV \\
Fe$_{\rm broad}$ Eq. width (eV)&	$93\pm13$		&	$82^{+19}_{-21}$		\\
$\chi^2/$d.o.f		&	$488/381$				&	$513/473$				\\
\enddata
\end{deluxetable}

\subsection{Spectral modeling}\label{sec:mcg_spec}
To interpret the shape of the lag-energy plot, we turn to spectroscopy. The spectrum of \mcg above 2 keV has been studied extensively. It was established that it has a broad iron line component, but not as extreme as some other objects. \cite{2007PASJ...59S.301R} for example, using Suzaku and XMM data, measure a line width of 62 eV, and an inner accretion disk radius of $37^{+25}_{-10}$ \rg when fitted with a diskline model (\citealt{1989MNRAS.238..729F}). If a full reflection model is used, the inner radius is $26^{+35}_{-8}$ \rg if the emissivity index is constrained to be $q=3$ (emissivity $\propto r^{-q}$), and goes to $\sim 6$\rg if $q$ is allowed to vary in the fit. Our fit for the XMM data gives 90\% confidence upper limit on the inner radius of 18 \rg and a best fit of 5.6 \rg and a relatively flat emissivity index $q=1.8\pm0.2$. We used XMM spectrum only to remain consistent with the timing analysis.

\begin{figure}[ht]
\includegraphics[width=0.46\textwidth,clip]{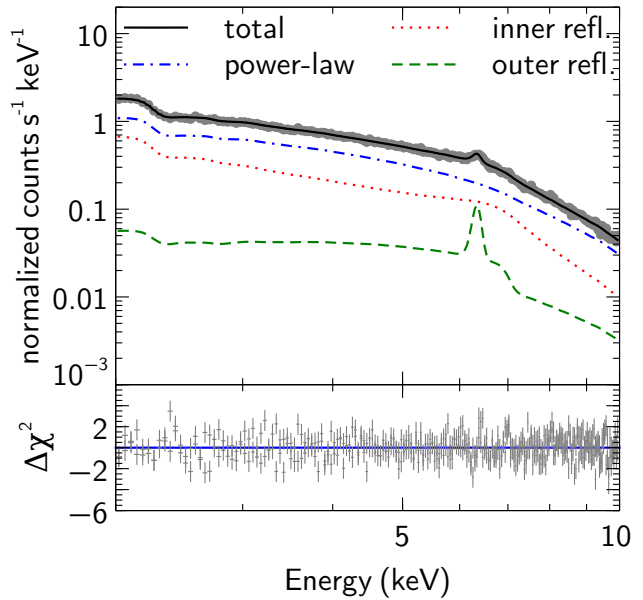}
\caption{The best fitting spectral model for \mcg. The model includes a power-law component (blue dash-dot line), outer reflection (green dashed line) modeled with \texttt{reflionx} and a relativistically-broadened inner reflection (red dotted line), modeled with \texttt{kdblur*reflionx}}
\label{fig:mcg_spec}
\end{figure}

Fig. \ref{fig:mcg_spec} shows the model used to fit the data between $2-10$ keV, and the best fitting parameters are presented in table \ref{tab:fits}. There are two reflection components modeled using the reflection table \texttt{reflionx} (\citealt{2005MNRAS.358..211R}): The outer component has a characteristic narrow iron line and originates at large distances from the black hole (molecular torus or broad line region) and is not expected to produce any variability at the time-scale of $\sim 10$ ks probed by the lag analysis in sec. \ref{sec:mcg_lag}. The inner reflection component, modeled with the reflection table \texttt{reflionx} broadened by a relativistic kernel \texttt{kdblur}, is more likely to be responsible for the lags by responding to variability in the direct power-law component.

Because both power-law and reflection contribute to every band between $2-10$ keV, and if we assume, to a first order, the delay between the two components to be energy-{\it independent}, then each point in the lag plot in Fig. \ref{fig:mcg_lagen2} gives a measure of the reflection fraction at that energy (variable reflection/power-law) multiplied by the intrinsic lag between the power-law and the reflection components (see further discussion in sec. \ref{sec:disc}.)

\section{NGC 7314}\label{sec:ngc}
\subsection{Time lags}\label{sec:ngc_lag}
\begin{figure}[ht]
\includegraphics[width=0.36\textwidth,clip]{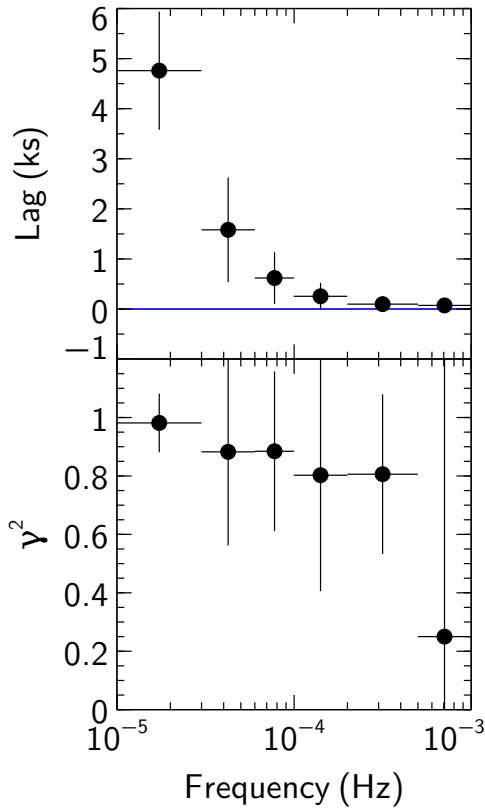}
\caption{The lag (top) and coherence (bottom) as a function of Fourier frequency between $4-5$ and $6-7$ keV energy bands for \ngc.}
\label{fig:ngc_lag}
\end{figure}

Fig. \ref{fig:ngc_lag}-top shows the frequency-dependent lags for \ngc between the the $4-5$ and $6-7$ bands. Similar to \mcg , there is a hard positive lag, where the peak of the iron K line lags the $4-5$ energy band, with a lag that depends on frequency. Fig. \ref{fig:ngc_lag}-bottom shows the measured coherence function (see sec. \ref{sec:mcg_lag}). The two bands have a relatively high coherence of $\sim 0.8$ up to $\sim5\times10^{-4}$ Hz.

\begin{figure}[ht]
\includegraphics[width=0.46\textwidth,clip]{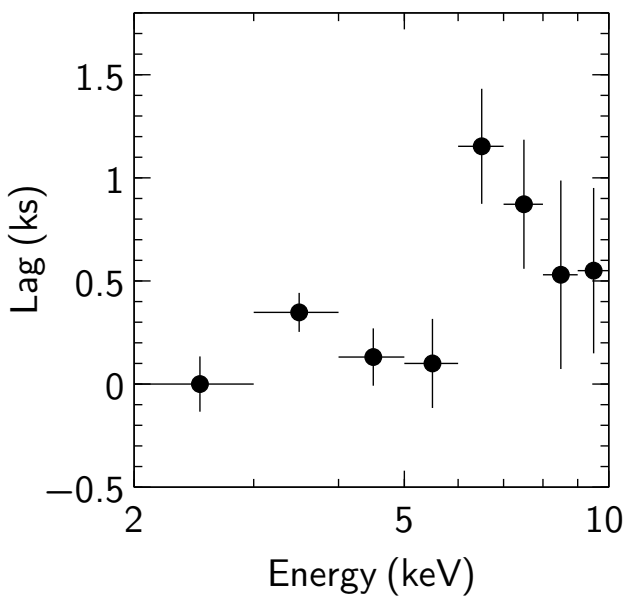}
\caption{Energy-dependent lag for \ngc at frequency of $(1-30)\times10^{-5}$ Hz. The whole $2-10$ keV band is used as a reference.}
\label{fig:ngc_lagen}
\end{figure}

Similar to the analysis in Sec. \ref{sec:mcg_lag}, we calculate the energy-dependent lags and show them in the Fig. \ref{fig:ngc_lagen}. there is a clear jump in the lag at $\sim5.5$ keV. Its significance is more than $99\%$. The line profile traced by the lag is narrower than that of \mcg and NGC  4151 (\citealt{2012MNRAS.422..129Z}). We also tested for frequency-dependence of the profile. The narrow feature at $6-7$ keV does not change shape if we select frequencies say $<5\times10^{-5}$ Hz. There is only a change in the magnitude of the lag, which can already be seen in Fig. \ref{fig:ngc_lag}-top. The constant shape with frequency is probably due to the sharpness of the jump ( or equivalently, the narrowness of the line ) making it difficult to detect a shape change.

\subsection{Spectral modeling}\label{sec:ngc_spec}
We also performed spectral fitting of the XMM observations. Previous spectral decomposition based on variability showed that the spectrum at $\sim 6$ keV contains a narrow peak and wings (\citealt{1996ApJ...470L..27Y}), indicating that it possibly contains a broad component. Using the high resolution capabilities of Chandra, \cite{2003ApJ...596...85Y} showed that the spectrum also contains emission from ions up to Fe \textsc{xvii} at $\sim$ 6.4 keV, as well as transient emission features from high ionization species Fe \textsc{xxv} and Fe \textsc{xxvi} Ly$\alpha$. All these lines appear to be variable and respond to continuum changes on in less than 12 ks. 
Although it would be interesting to attribute the lags in Fig. \ref{fig:ngc_lagen} to a particular line or ionization state, the energy resolution of the lags permitted by the current data (without significantly affecting the signal) cannot achieve that, and remains to be explored with future data.

\begin{figure}[t]
\includegraphics[width=0.46\textwidth,clip]{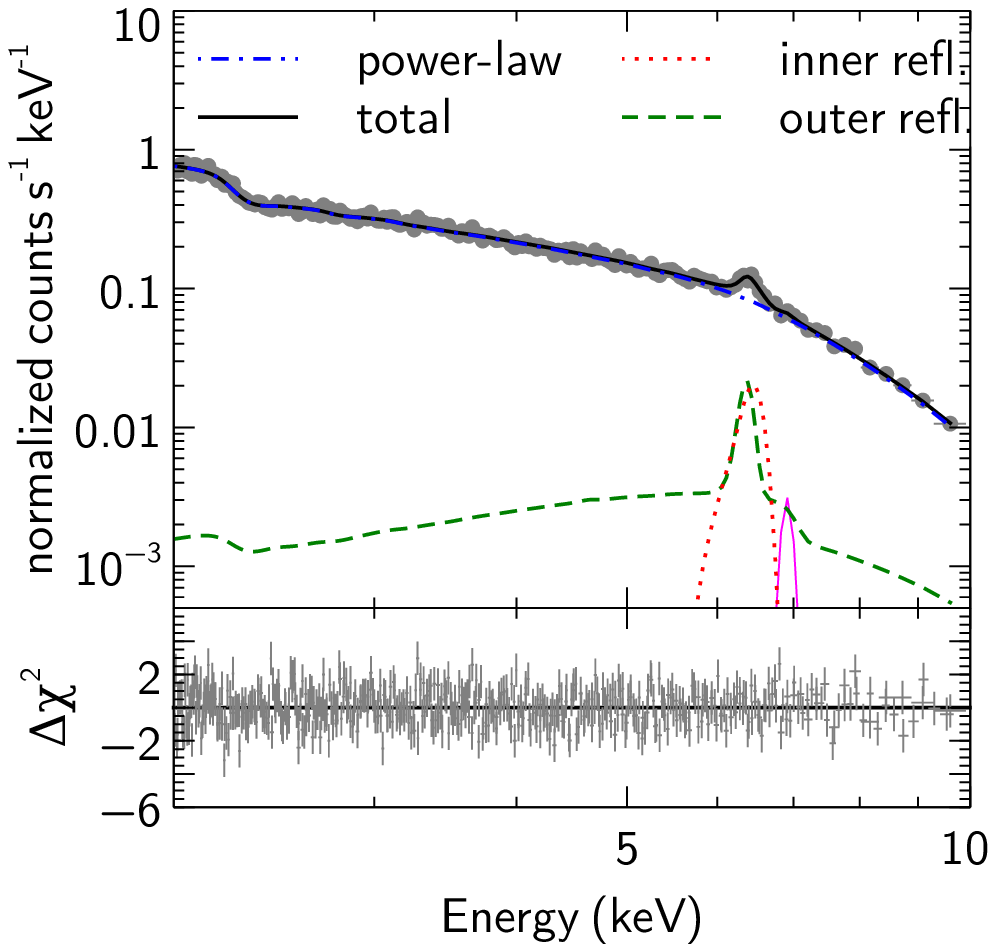}
\caption{The best fitting spectral model for \ngc. The model includes a power-law component (blue dash-dot line), outer reflection (green dashed line) modeled with \texttt{reflionx} (setting the ionization parameter to the lowest possible value of 1) and an inner reflection (red dotted line), modeled with \texttt{laor} and a narrow Fe \textsc{xxvi} line.}
\label{fig:ngc_spec}
\end{figure}

The spectrum from the XMM observation was analyzed. There is a clear line at $\sim 6.4$ keV. If fitted with a gaussian, it has an energy of $6.43\pm0.02$ keV and a width of $126\pm35$ eV, slightly resolved. To model it properly, we included the partially ionized reflection table model \texttt{reflionx} in addition to an absorbed power-law. The residuals show the presence of a line at $\sim 6.9$ keV, which was modeled using a narrow gaussian. This could be due to Fe \textsc{xxvi} (see \citealt{2011A&A...535A..62E} for full spectrum analysis). There were also residuals at the wings of the 6.4 keV line that are not accounted for by the cold reflection model. This could be due to a broad component (\citealt{1996ApJ...470L..27Y}) and we model them with a gaussian or a broad relativisitc line \texttt{laor} (\citealt{1991ApJ...376...90L}).

Although the broad component is not very broad, its presence is very significant ($>99.999\%$ confidence when fitted with a gaussian). The inner radius of the emitting region from the \texttt{laor} model is not very constrained, but it extends down to 13 \rg with an optimum value of 24 \rg, with nearly face-on viewing angle of $\theta=9^{\circ}$ and a flat emissivity of $q=2$, similar to the results found by \cite{2003ApJ...596...85Y} using Chandra observations and different arguments. The best fitting model is shown in Fig. \ref{fig:ngc_spec}, and the best fitting parameters are presented in table \ref{tab:fits}. It should be noted that, although these are the optimum parameters, their uncertainties are relatively large.
If more sophisticated models are used (e.g. \texttt{kdblur*reflionx}), we obtain parameters that are similar to those obtained from using the \texttt{laor} model.

Although it is not our aim to do detailed spectral modeling, it is very clear that the spectroscopy is very consistent with lag measurements made in Sec. \ref{sec:ngc_lag} (Fig. \ref{fig:ngc_lagen}), where there is a narrow line-like feature in the lag-energy plot. As we discussed in Sec. \ref{sec:mcg_lag}, the lag-energy plot, under simple assumptions, gives a measure of the reflection fraction as a function of energy (the variable reflection component responding to the direct power-law variations), and a reflection fraction that has a line peaking at $\sim 6.4$ keV is very consistent with the spectral modeling.

\section{Discussion}\label{sec:disc}
We have reported on the discovery of X-ray reverberation in two bright AGN in the iron K band. The analysis in Sections \ref{sec:mcg_lag} and \ref{sec:ngc_lag} present \textit{model-independent} measurements of time lags between the $6-7$ band where the Fe K$\alpha$ line peaks, and other bands, in a sense that this band always lags bands above and below. It is very apparent that these lags are linked to the emission in the Fe K$\alpha$ band and very likely are due to a reflection process.

The measurement of lag as a function of energy provides a way of measuring the reflection fraction as a function of energy in a completely independent way from the standard spectrum. This provides an invaluable tool to break the degeneracy often encountered in spectroscopic modeling, where it is often difficult to identify an underlying continuum due to absorption and emission complexities. To see how the lag-energy plot is measuring the reflection fraction, consider the following (see \citealt{2001AdSpR..28..267P,2001MNRAS.327..799K}): Let $S_{\rm d}(E,t)$ and $S_{\rm ref}(E,t)$ be the direct and reflected time-dependent spectra respectively, so that the total observed spectrum is:
\begin{equation}
S(E,t)=S_{\rm d}(E,t)+S_{\rm ref}(E,t)
\end{equation}
Assuming the direct component changes only in intensity and not in shape, we can write:
\begin{equation}
S_{\rm d}(E,t) = D(E) A(t)
\end{equation}
and the reflection can then be written as:
\begin{equation}
S_{\rm ref}(E,t) = f R(E) \int{A(t-\tau)T(\tau)d\tau}
\end{equation}
where $D(E), R(E)$ are the direct and reflected spectra respectively, $A(t)$ is the normalization of the direct component, $T(\tau)$ is the transfer function of the reflected component and $f$ is a constant that contain information about the geometry of the system. We have assumed again that $R(E)$ only changes in amplitude as it varies. The frequency-resolved phase lag as a function of energy is then:
\begin{equation}
tan(\phi(E,f)) = \frac{Im[\hat{S}^{\ast}(E,f) \hat{D}(E,f)]}{Re[\hat{S}^{\ast}(E,f) \hat{D}(E,f)]}
\end{equation}
where $\hat{}$ denotes the Fourier transform and $\ast$ denotes the complex conjugate. $Re$ and $Im$ indicate real and imaginary parts of a complex quantity respectively. This, for small reflection fractions then reduces to:
\begin{equation}\label{lag_eq}
tan(\phi(E,f)) = f \frac{R(E)}{D(E)} Im[\hat{T}(f)]
\end{equation}
Therefore for a single frequency band, the lag as a function of energy ( e.g. Fig. \ref{fig:mcg_lagen} and \ref{fig:ngc_lagen}) gives a measure of the reflection fraction as a function of energy.

Could these lags be due to the narrow Fe K$\alpha$ component? Narrow Fe K$\alpha$ lines are very common in AGN spectra. Although their origin is still not very clear (molecular torus, the broad line region or the outer parts of the disk itself, e.g. \citealt{2006MNRAS.368L..62N}), it is known that the narrow component of the line is not very variable generally, and certainly not on the time-scales probed by a typical single observation ($\sim 100$ ks) such as those presented here (e.g. \citealt{2004MNRAS.348.1415V, 2010MNRAS.408.1020B}). For the data used in this work, we calculated the RMS spectra to assess the level of variability as a function of energy. We found that there is a drop in the RMS at $\sim 6$ keV. The width of the trough cannot however be constrained by the data, and cannot therefore be used to confirm or rule out the origin of variability in the narrow component. 

There are however other lines of evidence against the observed lags being due to the narrow Fe K$\alpha$ line. The lags presented here, and that of NGC 4151 (\citealt{2012MNRAS.422..129Z}) appear in energy-dependence to be a continuation of the lags seen in the soft excess in many Narrow Line Seyfert 1 galaxies. This is particularly apparent in the lag-energy dependence of 1H0707-495 (\citealt{2011MNRAS.412...59Z, 2013MNRAS.428.2795K}), RE J1034+396 (\citealt{2011MNRAS.418.2642Z}) and IRAS 13224 (Kara et al.). Also, as in the case of NGC 4151, the observed peak in the lag-energy spectrum is less than 6.4 keV. If it was produced by the narrow component, the peak would be at the rest frame energy of the Fe line. Furthermore, the fact that the lag-energy shape depends on time-scale (i.e. temporal frequency) argues again against an origin in the narrow component.

Now we consider the magnitude of the lags. The light-crossing time around a black hole at a radius $r$ (in units of  $r_g = GM/c^2$) is $5M_{6}r$ seconds, where $M_6$ is the black hole mass in units of $10^{6}$ M$_{\odot}$. For $M_{6} = [50,5]$, where the first number is for \mcg  and second for \ngc, the measured inner radii of emission inferred from the spectral fitting of $r=[5.6,24]$ (in \rg) give a light crossing time of $[1.4,0.6]$ ks, within a factor of 2 of the measured delays in Fig.\ref{fig:mcg_lagen2} and \ref{fig:ngc_lagen}. It should be noted however that associating the measured lag directly with a particular radius is not straight forward. There are many factors that need to be considered. First, the measured lags in Fig. \ref{fig:mcg_lagen} and \ref{fig:ngc_lagen} include a dilution factor (equation \ref{lag_eq}) that includes the reflection fraction which makes the measured lags between bands {\it smaller} than the intrinsic lag between the direct and reflected components. On the other hand, the effects of geometry and gravitational Shapiro delays make the radii inferred from light-crossing estimates {\it larger} than they are. For example, if the illuminating source is high above the disk, which could be the case for \ngc, then the delay is larger than the light-crossing time at the radius of emission inferred from spectral fitting. Accounting for these effects requires full modeling that is beyond the scope of this work (see e.g. \citealt{2012arXiv1212.2213W}) and would require more data to study the lag-energy in finer energy bins and frequency bands.

Furthermore, as Fig. \ref{fig:mcg_lagcoh} and \ref{fig:ngc_lag} show, the lag is frequency-dependent and a single measured lag at a single frequency cannot be trivially linked to a particular radius. It is however tempting to broadly associate small lags at small time-scales with small radii and longer lags with larger radii, The reason is that the competing effects of dilution vs geometry and gravitational delays roughly cancel out (though not exactly, see  \citealt{2012arXiv1212.2213W}). Therefore, the increase in lag magnitude seen in Fig. \ref{fig:mcg_lagen} with decreasing Fourier frequency is likely caused by small time-scale variations at small radii (and hence small lags) being filtered out. The effect is similar to, and strongly supported by, the time-scale-dependent shape of the lag-energy spectrum of NGC 4151 (\citealt{2012MNRAS.422..129Z}). This highlights the power of these plots that measure the reflection fraction independently of the classical spectrum, allowing us to use our knowledge of how the reflection fraction changes with time-scales to interpret the lags and the spectra.

It is also interesting to examine the time-scale axis of the variability (i.e. Fourier frequency axis). Although the variability in black holes is well established by observations, it is not clear what physical time-scale in the accretion disk it represents. For example, if we consider the viscous time-scales $t_{\nu}=5r^{1.5}\alpha^{-1}(H/R)^{-2}M_6$ seconds, where $r$ and $M_6$ are again in units of \rg and $10^{6}$ M$_{\odot}$ respectively, and $\alpha$ is the disk viscosity parameter (\citealt{2002apa..book.....F}), then the measured radii from spectral fitting of $[5.6,24]$ \rg give a viscous frequency $1/t_{\nu}$ of $[7\times10^{-8},8\times10^{-8}]$ Hz, using $\alpha=0.1$ and $(H/R)=0.05$. The measured lags clearly span higher frequencies (shorter time-scales) than this. If instead thermal time-scales are considered, where $t_{th}=(H/R)^2t_{\nu}$, then the measured radii give frequencies of $\sim3\times10^{-5}$ Hz in both cases, which falls exactly in the observed range over which the reverberation lag is measured. Thermal time-scales could dominate if variability in density and/or ionization are important. Thermal time-scale seems to also be most relevant in the case of accreting white dwarfs too (e.g. \citealt{2012arXiv1208.6292S} for some recent results).

In summary, we have presented lag measurements in the Fe K band that show the peak of the line $6-7$ keV band lagging bands either side of it. These simple measurements are model-independent, and to interpret them, we showed that they are consistent with a broad iron line responding to variations in an illuminating continuum. Most of the emission originates at 5 \rg for \mcg and 24 \rg for \ngc. These two objects are selected for their brightness and high variability, two other variable NLS1 appear to show similar iron K lags (\citealt{2013MNRAS.428.2795K,2013MNRAS.tmp..654K}), which is a possible indication that these lags are common among variable objects. The question of how common are they among radio quiet AGN in general remains however to be explored.

\bibliography{main}

\end{document}